\documentclass[11pt]{article}
\usepackage{setspace}
\usepackage{amssymb,amsfonts,amsmath}
\usepackage{cite,enumerate,float,comment}
\usepackage{color}
\usepackage{tikz}
\usetikzlibrary{arrows.meta}
\usetikzlibrary{backgrounds}
\usepackage{xcolor}
\definecolor{mintcream}{rgb}{0.96, 1.0, 0.98}
\definecolor{champagne}{rgb}{0.97, 0.91, 0.81}
\definecolor{bubblegum}{rgb}{0.99, 0.76, 0.8}
\definecolor{airforceblue}{rgb}{0.36, 0.54, 0.66}
\definecolor{persianblue}{rgb}{0.11, 0.22, 0.73}
\definecolor{zaffre}{rgb}{0.0, 0.08, 0.66}
\usepackage[pdftex,unicode,
colorlinks=true,
linkcolor = zaffre,
citecolor = cyan,
urlcolor = airforceblue]{hyperref}
\usepackage{cleveref}
\usepackage{tcolorbox}

\usetikzlibrary{arrows,snakes,backgrounds}

\def\be{\begin{eqnarray}}
\def\ee{\end{eqnarray}}
\def\nn{\nonumber}

\def\p{\partial}

\definecolor{red}{rgb}{1,0,0}
\definecolor{orange}{rgb}{1,0.5,0}
\definecolor{violet}{rgb}{0.7,0,1}

\newcommand{\EE}{\hat{\mathcal{E}}}

\usepackage[top = 2 cm, bottom = 2 cm, left = 2.5 cm, right = 1.5 cm]{geometry}

\usepackage[utf8]{inputenc}

\begin{document}

\hfill MITP-TH-18/24

\hfill ITEP-TH-24/24

\hfill IITP-TH-19/24

\vspace{1cm}
\centerline{\LARGE{
	Position space equations for generic Feynman graphs
}}

\bigskip
\vspace{0.5cm}

\centerline{{\bf V.Mishnyakov $^{a,}$\footnote{victor.mishnyakov@su.se}, 	A. Morozov$^{b,c,d,}$\footnote{morozov@itep.ru},	M.Reva$^{b,c,}$\footnote{reva.ma@phystech.edu}} }

\bigskip
\vspace{0.2cm}

\begin{center}
	$^a$ {\small {\it Nordita, KTH Royal Institute of Technology and Stockholm University,}}\\{\small {\it 
			Hannes Alfv\'ens v\"ag 12, SE-106 91 Stockholm, Sweden}}
	\\
	$^b$ {\small {\it MIPT, Dolgoprudny, 141701, Russia}}\\
	$^c$ {\small {\it NRC ``Kurchatov Institute", 123182, Moscow, Russia}}\\
	$^d$ {\small {\it Institute for Information Transmission Problems, Moscow 127994, Russia}}\\
    $^e$ {\small {\it ITEP, Moscow, Russia}}\\
\end{center} 
\vspace{0.8cm}
\centerline{\bf \normalsize Abstract}
\vspace{0.8cm}

{
We propose the extension of the position space approach to Feynman integrals from the banana family to generic Feynman diagrams. Our approach is based on getting rid of integration in position space and then writing differential equations for the products of propagators defined for any graph. We employ the so-called ''bananization'' to start with simple Feynman graphs and further substituting each edge with a multiple one. We explain how the previously developed theory of banana diagrams
can be used to describe what happens to the differential equations
(Ward identities) on Feynman diagrams after this transformation. Our approach works for generic enough (large enough) dimension and masses. We expect that after Fourier transform our equations should be related to the Picard-Fuchs equations. Therefore, we describe the challenges of Fourier transform that arise in our approach.

	
}

\newpage



\section{Introduction}

Feynman integrals of perturbative QFT are known to be notoriously hard to calculate. In recent years, however, great progress has been achieved both on the quantitative and the qualitative fronts \cite{Weinzierl:2022eaz}.  A lot of algorithmically effective tools have been developed for computing specific Feynman integrals, leading, in particular, to leaps in precision collider phenomenology. At the same time, various mathematical properties have been established both generic and for special (families) of graphs. Among these are number theoretic properties \cite{Bonisch:2022mgw, Candelas:2019llw}, geometric \cite{Bonisch:2021yfw, Marcolli:2009zy, Doran:2023yzu}, Hopf-algebraic \cite{Gerasimov:2000pr, Kreimer:1997dp, Connes:1999yr}, integrable \cite{Loebbert:2022nfu} and others. Among others, an especially effective way to treat Feynman integrals is finding differential equations that they satisfy. These come in various shapes and forms, such as the systems coming from IPB identities \cite{Chetyrkin:1981qh, Lee:2008tj}, the Picard-Fuchs equations \cite{Lairez:2022zkj, delaCruz:2024xit, Mishnyakov:2024rmb}, GKZ-hypergeometric systems \cite{Vanhove:2018mto, delaCruz:2019skx}, and the recently introduced position space ones \cite{Mishnyakov:2024rmb, Mishnyakov:2023wpd,Cacciatori:2023tzp}. Only in some cases relations between these various kinds are well understood.

A common drawback of various differential equations approach is a lack of generic statements, that would be valid for any graph. In principle even the existence of these equations is not completely clear, and deriving them involves  quite a lot of non-generic choices. 
Despite that there is progress on in that direction too. 

We continue the development of the position space approach. In previous works \cite{Mishnyakov:2022uer,Mishnyakov:2023sly,Mishnyakov:2023wpd,Mishnyakov:2024rmb} we were focused mostly on the so-called banana graphs, which are a certain family of multi-loop two-point functions. In the latest work we have hinted a possible generalization to other graphs. Here we would like to take a step further and propose that the position space approach allows demonstrating from very simple first principles that \textbf{differential equations exist for generic Feynman integrals.}

 Our proposal relies heavily on the idea of considering what we think is appropriate to be called \emph{unintegrated correlators}, even though in practice we deal with single diagram contributions to correlators. As we explain below, this means that we consider position space expression which are products of propagators which do not have any integration over the would-be internal vertices. These functions have two key properties. They satisfy conceptually very simple differential equations (which might be quite involved technically) and in momentum space they correspond to a function, whose various evaluations correspond to different choices of integrated vertices.

\section{The setting}

Consider a graph $\Gamma$, and let it be a \emph{simple graph} for now, meaning that it does not have multiple edges between any vertices. Denote $V_\Gamma,E_\Gamma$ the set of vertices and edges. The corresponding ''Feynman integral'' which may be called the ''Feynman function'' is just a product of propagators over all edges:
\begin{equation}\label{eq:def1}
	I_\Gamma\left( \mathbf{x} | \mathbf{m}^2 ,D \right) =\prod_{<ij> \in E_\Gamma}G(x_{i}^\mu-x_j^\mu| m^2_{ij},D)
\end{equation}
where $\mathbf{x}$ is a collection of vertex coordinates $x_i^\mu \in \mathbb{R}^D$ are coordinates correspond to each vertex $i \in V_\Gamma$ and, in principle each propagator carries its own mass $m_{ij}^2$, collectively denoted by $\mathbf{m}^2$. Feynman diagrams can have multiple edges, which we could incorporate in the definition \eqref{eq:def1} straight away. However, for the purposes of writing equations we would like to think of starting with a simple graph and then multiplying each edge - a procedure we call ''bananization'' - meaning that we substitute each propagator with the corresponding banana function, which is just a product of propagators:
\begin{equation}
	B_n(x_i-x_j)= \prod_{a=1}^{n} G(x_{i}^\mu-x_j^\mu| m^2_{ij,a},D)
\end{equation}  
These functions $I_\Gamma\left( \mathbf{x} | \mathbf{m}^2 ,D \right)$ are what we refer to as ''unintegrated correlators''. This means that any true position space Feynman integral can be obtained by declaring some edges to be internal and integrating over them.
\\

To set further set up our story consider the Fourier transform of these functions:
\begin{equation}
	\begin{split}
		I_\Gamma\left( \mathbf{x} | \mathbf{m}^2 ,D \right) &= \int \prod_{<ij> \in E_\Gamma} d^Dp_{ij} \exp\left(\sum_{<kj> \in E_\Gamma} i p_{kj}\cdot (x_k-x_j) \right) \prod_{<ij> \in E_\Gamma} {B_n}(p_{ij})  = 
			\\
			& =\int \prod_{<ij> \in E_\Gamma} d^Dp_{ij}\exp\left(\sum_{k \in V_\Gamma} i  (x_1-x_k)\cdot \left( \sum_{j<k} p_{jk} - \sum_{j>k} p_{kj} \right) \right) \prod_{<ij> \in E_\Gamma} {B_n}(p_{ij})
	\end{split}
\end{equation} 
where by $p_{ij}$ we mean all momenta from $i$-th to $j$-th vertex. After a partial change of variables we separate the external from the loop momenta:
\begin{equation}
	\begin{split}
	&k_i= \left( \sum_{j<k} p_{jk} - \sum_{j>k} p_{kj} \right) \,  \, i =1,\ldots |V_\Gamma| - 1 \ ,\\
	&l_{i}= \text{orthogonal complement to $k_i$} 
\end{split}
\end{equation}
and obtain 
\begin{equation}\label{eq:Fourier3}
	I_\Gamma\left( \mathbf{x} | \mathbf{m}^2 ,D \right) =\int \prod_{i=1}^{|V_\Gamma|} d^D k_{i}\exp\left(\sum_{j \in V_\Gamma} i  (x_1-x_j)\cdot k_j \right) I_\Gamma(
	\mathbf{k} )
\end{equation}
where:
\begin{equation}
	I_\Gamma( 	\mathbf{k})  = \int \prod_{i=1}^{L_\Gamma} d^D l_i \prod_{j=1}^{|E_\Gamma|} B_{n_j}\left(\{l,k\}  \right)
\end{equation}
is a loop integral. The function $I_\Gamma( 	\mathbf{k})$ depends on the same number of independent variables as the coordinate representation, i.e. $\frac{|V_\Gamma|(|V_\Gamma|-1)}{2}$ squared momenta and their scalar products for sufficient big space-time dimension $D$. We will elaborate this statement in the next section. Any ''true'' Feynman integral corresponds to declaring a subset of vertices to be integral. In momentum space this means that one has to set all the corresponding momenta to zero $k_i =0$ for $i \in V^{\text(int)}_\Gamma$. In this sense $I_\Gamma( 	\mathbf{k})$ it is a more general function than any of the particular loop integrals. 

\section{On the choice of independent variables}\label{sec:ChoiceOfVariables}

As usual, the Feynman functions do not depend on all the position variable $x_i^\mu$, but on their Poincare invariants. The most obvious choice of invariants is just the lengths:
\begin{equation}
    X_{ij}=\sqrt{(x_i^\mu-x_j^\mu)(x_{i,\mu}-x_{j,\mu})} 
\end{equation}
not for just the edges, but for all pairs of coordinates. Note, that we use upper case for invariants and lower case for the vectors.  Hence the function depends on $\dfrac{|V_\Gamma|(|V_\Gamma|-1)}{2}$ variables. This is true if all the variables are independent, which requires for the dimension $D$ to be big enough. 
It's easy to see that in $D=1$ for, say $|V_\Gamma|=3$ these variables are not independent. For $D=2$ they are, however, for $|V_\Gamma|=4$ one has to at least $D=3$ for the length of side of the tetrahedron to be independent variables. In general, the variables $X_{ij}^2$ are independent if the following condition is satisfied:
\begin{equation}
	D \geq |V_\Gamma|-1 \, .
\end{equation}  

The synopsis of the next section is to show that coordinate space equations are straightforward in terms of the invariants. However, this one could argue that these variables are not completely natural. \emph{A priory} the Feynman function is formulated as the function of vertex position. In particular, the Fourier transform utilizes such presentation. Indeed, as we later discuss in  section \ref{sec:Fourier} writing the equation in terms of the invariants  causes the Fourier transform to be complicated, even though, in principle, possible. Moreover, when written in terms of invariants, $x_{ij}^2$ the topology of the graphs is somewhat lost.
\\

To demonstrate that, we consider two graphs:
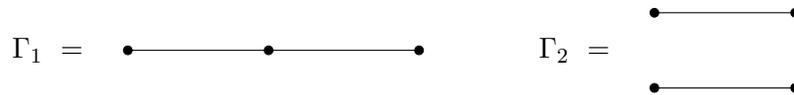
\vspace{0.5cm}
\begin{center}
    \begin{tikzpicture}
   \node (A) at (-1,0) {$\Gamma_1 \ =$}; \draw[{Circle}-{Circle}] (0,0) -- (2,0);
    \draw[-{Circle}] (2,0) -- (4,0);
     \node (B) at (6,0) {$\Gamma_2 \ =$}; \draw[{Circle}-{Circle}] (7,0.5) -- (9,0.5);
    \draw[{Circle}-{Circle}] (7,-0.5) -- (9,-0.5);
\end{tikzpicture}
\end{center}
\vspace{0.5cm}
These two graphs corresponding to similar functions:
\begin{equation}
	I_{\Gamma_1} = G(X^2_{12})G(X^2_{23}) \, , \quad I_{\Gamma_{2}}= G(X_{12}^2)G(X_{34}^2)
 \label{example_graphs}
\end{equation}
As functions of invariants these are the same functions: product of two propagators each depending on its own variable. Therefore if we want to actually distinguish between the two and keep track of the topology we should keep some part of the data from the initial variables $x_i^\mu$. Mainly we should remember, that $X_{ij}^2=(x_i-x_j)^\mu (x_i-x_j)\mu$. This actually means, that we know that the function $I_{\Gamma_1}$ actually involves three points and therefore generally depends on the variable $X_{13}^2$ as well, just that the dependence is constant. Similarly we remember that $I_{\Gamma_2}$ is a function of four vertex positions and therefore depends (though constantly) on $X_{13}^2,X_{14}^2, X_{23}^2$ as well. We will see how to take this into account in the next section.

\section{Differential equation}\label{sec:Fourier}

In this section we demonstrate the position space equations that the unintegrated correlators satisfy. A review of the banana case is available in the appendix \ref{sec:Appendix1}.

\subsection{Generic graphs and bananization}

Let us first start with the simple graph again (before bananization).
 Each propagator satisfies the (Klein-Gordon) equation of motion, which we write in $\Lambda$-form for convenience:
 \begin{equation}
 	\left(\Lambda_{ij}^2 +(D-2)\Lambda+X_{ij}^2m_{ij}^2 \right) G\left(X_{ij}^2\Big|m_{ij},D \right) =0 
  \label{KG_eq}
 \end{equation}
where $\Lambda_{ij}=X_{ij}\dfrac{\partial}{\partial X_{ij}}$ just as before 
\begin{equation}
	X_{ij}^2=(x_i-x_j)^\mu (x_i-x_j)_\mu
\end{equation}
We also denote the respective differential operator as:
\begin{equation}
	\hat{\mathcal{E}}_{1}\left(X_{ij} \right) := \left(\Lambda_{ij}^2 +(D-2)\Lambda+X_{ij}^2m_{ij}^2 \right)
\end{equation}
as it is first in the series of banana graph operators, introduced in previous works \cite{Mishnyakov:2023wpd, Mishnyakov:2023sly, Mishnyakov:2024rmb}. Therefore the Feynman function $I_\Gamma$ satisfies a system of equations
\begin{equation}
	\EE_1(X_{ij}) I_\Gamma(\mathbf{x}) = 0 \, ,\quad  <ij> \in E_\Gamma
\end{equation}
Promoting to bananized diagrams we substitute each edge by a banana graph and, hence, the resulting differential equation for each line is the respective banana operator. The corresponding operators is constructed by taking the symmetric product of the equations of motion for a single propagator \eqref{KG_eq} via an explicit procedure nicknamed the $\Lambda$-formalism \cite{Mishnyakov:2023sly} and reviewed in \ref{sec:Appendix1}. Therefore for the banananized graph, where the $i \leftrightarrow j$ edge has been substituted by and $n_{ij}$-multiple edge one has the following system:
\begin{equation}
\EE_{n_{ij}}(X_{ij}) I_\Gamma(\mathbf{x}) = 0 \, ,\quad  <ij> \in E_\Gamma
\end{equation}
\\
To really specify the function we also need to tell which of the pairs $<i,j>$ do not belong to the set of edges, meaning that the power of the propagator is vanishing. This can be actually also nicely incorporated in the whole scheme, just by declaring a constant function to be a 0-banana function, and writing the corresponding equation as:
\begin{equation}
	\hat{\mathcal{E}}_{0}\left(X_{ij} \right) B_0:= \Lambda_{ij} B_0 = 0 \,  \leftrightarrow \, B_0 = \text{const}
\end{equation}
Therefore the full system of equations is given by:
\begin{tcolorbox}[colback=champagne!30]
	\begin{equation}
		\EE_{n_{ij}}(X_{ij}) I_\Gamma(x) = 0 \, , i,j \in V_\Gamma
	\end{equation}
\end{tcolorbox}
where $n_{ij}$ is the number of edges between two vertices, including zero's if the vertices are not connected. 
\\

Finally, it is crucial that we remember that the invariants are indeed made out of the initial coordinates, which will further be important in Fourier transform. This means that we want to distinguish between say a triangle graph and three independent lines. Specifying the set of vertices and including the 0-banana  operators is important for that. Consider the two graphs from \eqref{example_graphs}:
\begin{equation}
	I_{\Gamma_1} = G(X^2_{12})G(X^2_{23}) \, , \quad I_{\Gamma_{2}}= G(X_{12}^2)G(X_{34}^2)
\end{equation}
However to truly differentiate between them we should treat the first as a function of three vertex coordinates and the second as a function of 4 vertex coordinates with the following equations:
\begin{equation}
  \EE_1(X_{12})	I_{\Gamma_1}(\mathbf{x}) =0 ,\quad  \EE_1(X_{23})	I_{\Gamma_1}(\mathbf{x}) =0, \quad  \EE_0(X_{13})	I_{\Gamma_1}(\mathbf{x}) =0
\end{equation}
while for the second graph:
\begin{equation}
	\begin{split}
			 &\EE_1(X_{12})	I_{\Gamma_2}(\mathbf{x}) =0 \quad  \EE_1(X_{34})	I_{\Gamma_2}(\mathbf{x}) =0
			 \\
			 &\EE_{0}(X_{1i})	I_{\Gamma_2}(\mathbf{x}) =0,  \ i =2,\ldots 4 \quad \EE_{0}(X_{2i})	I_{\Gamma_2}(\mathbf{x}) =0,  \  i =3,4
	\end{split}
\end{equation}

\subsection{Examples in the massless case}

When the propagators are massless, the position space equations are the simplest and can be given explicitly. The derivation and notations are explained in appendix \ref{sec:Appendix1}. For a single banana graph, the annihilating differential operator, denoted by $\widehat{\rm DET}_n$ is given by:
\begin{equation}
    \widehat{\rm DET}_n =   \hat\Lambda \prod_{j=1}^n  \Big(\hat\Lambda+j(D-2)\Big)
\end{equation}
Therefore, say for the ice-cone, graph, which is a particular case of a bananized triangle (with only one multiple edge) one has:
\begin{equation}
    I_{\mathrm{Triangle}}(x_{1},x_2,x_3)=G(X_{12})G(X_{23})G(X_{13})^{n}
\end{equation}
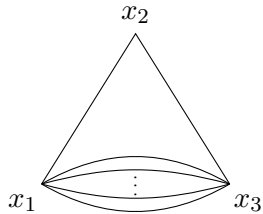
\vspace{0.5cm}
\begin{center}
    \begin{tikzpicture}
  \draw (0,0) -- (1.25,2) -- (2.5,0);
  \draw (0,0) to[out=15,in=180-15]  (2.5,0);
  \draw (0,0) to[out=30,in=180-30]  (2.5,0);
\draw (0,0) to[out=-15,in=180+15] (2.5,0);
    \draw (0,0) to[out=-30,in=180+30]  (2.5,0);
    \node (x1) at  (0-0.25,0-0.25) {$x_1$};
    \node (x2) at (1.25,2+0.25) {$x_2$};
    \node (x3) at (2.5+0.25,0-0.25) {$x_3$};
   \node (ldots) at (1.25,0) {\rotatebox{90}{\scalebox{0.65}{\dots}} } ;
\end{tikzpicture}
\end{center}
\vspace{0.5cm}
The equations are:
\begin{equation}
\begin{gathered}
         \hat\Lambda_{12} \Big(\hat\Lambda_{12}+(D-2)\Big)  I_{\mathrm{Triangle}}(x_{1},x_2,x_3) =  \hat\Lambda_{23} \Big(\hat\Lambda_{23}+(D-2)\Big)  I_{\mathrm{Triangle}}(x_{1},x_2,x_3) = 0
         \\
         \hat\Lambda_{13} \prod_{j=1}^n  \Big(\hat\Lambda_{13}+j(D-2)\Big)I_{\mathrm{Triangle}}(x_{1},x_2,x_3) = 0
\end{gathered}
\end{equation}

It is also instructive to consider the four point case, where one could have more interesting topologies. Consider, for example, a bananized kite graph:
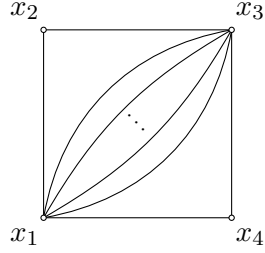
\vspace{0.5cm}
\begin{center}
    \begin{tikzpicture}
  \draw (0,0) -- (2.5,0) -- (2.5,2.5) -- (0,2.5) -- (0,0);
  \draw (0,0) to[out=10,in=-100] (2.5,2.5);
   \draw (0,0) to[out=30,in=-120] (2.5,2.5);
    \draw (0,0) to[out=90-10,in=-180+10] (2.5,2.5);
    \draw (0,0) to[out=90-30,in=-180+30] (2.5,2.5);
    \node (x1) at (-0.25,-0.25) {$x_1$};
    \node (x2) at (0-0.25,2.5+0.25) {$x_2$};
    \node (x3) at (2.5+0.25,2.5+0.25) {$x_3$};
    \node (x4) at (2.5+0.25,0-0.25) {$x_4$};
    \node (ldots) at (1.25,1.25) {\rotatebox{-45}{\scalebox{0.75}{\dots}} } ;
    \filldraw[color=black,fill=white] (0,0) circle (1pt);
     \filldraw[color=black,fill=white] (2.5,2.5) circle (1pt);
      \filldraw[color=black,fill=white] (2.5,0) circle (1pt);
       \filldraw[color=black,fill=white] (0,2.5) circle (1pt);
\end{tikzpicture}
\end{center}
\vspace{0.5cm}
\begin{equation}
    I_{\Gamma}(x_1,x_2,x_3,x_4) = G(X_{12})G(X_{23})G(X_{34})G(X_{14}) G(X_{13})^{n}
\end{equation}
This function satisfies a system of equation:
\begin{equation}
    \begin{gathered}
        \hat\Lambda_{12} \Big(\hat\Lambda_{12}+(D-2)\Big) I_{\Gamma}(x_1,x_2,x_3,x_4) = \hat\Lambda_{23} \Big(\hat\Lambda_{23}+(D-2)\Big) I_{\Gamma}(x_1,x_2,x_3,x_4) =0 \,,
        \\
        \hat\Lambda_{34} \Big(\hat\Lambda_{34}+(D-2)\Big) I_{\Gamma}(x_1,x_2,x_3,x_4) = \hat\Lambda_{14} \Big(\hat\Lambda_{14}+(D-2)\Big) I_{\Gamma}(x_1,x_2,x_3,x_4) =0\,,\\
        \hat\Lambda_{13} \prod_{j=1}^n  \Big(\hat\Lambda_{13}+j(D-2)\Big)I_{\Gamma}(x_1,x_2,x_3,x_4) =0\,,
        \\
        \hat\Lambda_{24}I_{\Gamma}(x_1,x_2,x_3,x_4) =0.
    \end{gathered}
\end{equation}
The last equation is necessary to fully specify the topology of the diagram — it ensures that there is no edge between the $1$ and $4$ vertex.
\\\\
Massless case was chosen only for simplicity reason and all the techniques in this article are applicable to case with masses too.

\section{Fourier transform}
One nice feature of the banana case is that by writing the very straightforward equation in position space and Fourier transforming it, one could obtain a momentum differential equation for the non-trivial loop integral. In the generic mass case the resulting equation seemed to be in a sense non-minimal (reducible), which, however, is a separate issue. Still it was demonstrated \cite{Mishnyakov:2023sly, Mishnyakov:2024rmb} that for the known examples  the rightmost factor of the differential operator is the irreducible Picard-Fuchs operator 
\\

Therefore it is natural to ask if it is possible to do so for the equations we presented for a generic graph. In this section, we demonstrate that this is indeed the case. Here we don't want to complicate things by discussing minimality. Moreover, we should concede that at the current stage, the resulting momentum differential operators are of completely intractable form. This phenomenon has already been visible for the triangle examples in \cite{Mishnyakov:2024rmb}.  However, in this paper we demonstrate a kind of proof of concept - that the equations always exist. The problem of choosing the correct variables/bases should be solved in the future.	  
\\

The main technical issue, which also lead to the equations being complicated is the fact the momenta couple to the positions $x^\mu_i$, while the equations are given in terms of the invariants $X_{ij}^2$. Moreover, due to symmetries, the momenta $k_i^\mu$ in \eqref{eq:Fourier3} couple to $(x_1-x_i)^\mu$.  Hence we have to find a set of differential operators that would on one hand reproduce $\Lambda_{ij}$ and $X_{ij}^2$, when acting on invariant function, and would be suitable for Fourier transform. 
\\

Naively one could attempt to write:
\begin{equation}
	\Lambda_{ij}= X_{ij} \dfrac{\partial}{\partial X_{ij} }  \quad \xrightarrow{?} \quad  (x_i-x_j)^\mu \dfrac{\partial^\mu}{\partial (x_i-x_j)^\mu}
\end{equation}
However, this is not correct, since the set of vectors  $(x_i-x_j)^\mu$ are not independent. Only their lengths are for suitably large $D$. Therefore an alternative solution should be found. One way to resolve this issue is to consider a set of operators:
\begin{equation}
	\mathcal{D}_{ab} = (x_a-x_b)^\mu \left(\dfrac{\partial^\mu}{\partial x_a^\mu}+\dfrac{\partial^\mu}{\partial x_b^\mu}\right)
\end{equation}
The plus sign between the derivatives makes the Fourier transformed operators simpler. After Fourier transform we get:
\begin{equation}
	\begin{split}
			(x_a-x_b)^\mu \ \   & \longrightarrow \ \ \dfrac{\partial}{\partial k_b^\mu} - \dfrac{\partial}{\partial k_a^\mu}
			\\
			\dfrac{\partial}{\partial x_a^\mu} \ \ \quad  &\longrightarrow \ \ -k_a^\mu
			\\
			\dfrac{\partial}{\partial x_1^\mu} \ \ \quad &\longrightarrow \ \ \sum_{a=2}^n k_a^\mu
			\\
			(x_1-x_a)^\mu \ \   &\longrightarrow  \ \ \dfrac{\partial^\mu}{\partial k_a^\mu}
	\end{split}
\end{equation}

Hence the $	\mathcal{D}_{ab}$ operators can be Fourier transformed. On the other hand, when acting on invariant functions we have the following equality.
\begin{equation}\label{eq:SystemLambdaToD}
	\begin{split}
			D_{ab}&= \sum_{i \neq a, b}(-X^2_{ib}+X^2_{ab}+X^2_{ia}) \dfrac{\partial}{\partial X_{ia}^2} -\sum_{j \neq a, b} (-X^2_{aj}+X^2_{ab}+X^2_{jb}) \dfrac{\partial}{\partial X_{jb}^2}= 
		\\
		&=\sum_{i \neq a, b}\dfrac{ (-X^2_{ib}+X^2_{ab}+X^2_{ia})}{2 X^2_{ia}} \cdot  \Lambda_{ia} -\sum_{j \neq a, b}\dfrac{ (-X^2_{aj}+X^2_{ab}+X^2_{jb})}{2X^2_{jb}} \cdot  \Lambda_{jb}
	\end{split}
\end{equation}
Recall that we want to express the set of operators $\Lambda_{ij}$ in terms of $D_{ab}$ and $x_{ij}^2$. The system of equations \eqref{eq:SystemLambdaToD} as a linear system on $\Lambda_{ij}$ almost does the job. A complication,  that requires a minor modification is due to the fact that this system is degenerate. It's easy to check that while the size of the system is $\frac{n(n-1)}{2}$ it has corank one. Despite that, the problem is easily resolved by substituting one of the equations in \eqref{eq:SystemLambdaToD} by:
\begin{equation}\label{eq:LambdaEquation}
	\Lambda_{\text{total}} = \sum_{i<j}\Lambda_{ij}
\end{equation}
where
\begin{equation}
	\Lambda_{\text{total}} : = \sum_{a=1}^n x_a^{\mu} \dfrac{\partial}{\partial x_a^\mu}
\end{equation}
is the total dilatation/Euler operator. It has a Fourier transform:
\begin{equation}
\sum_{a=1}^n x_a^{\mu} \dfrac{\partial}{\partial x_a^\mu} \ \ \longrightarrow \ \  - \sum_{a=2}^{n} \dfrac{\partial}{\partial k_a^\mu} k^\mu_a 
\end{equation}
The resulting system of equations, \eqref{eq:SystemLambdaToD} with one equation substituted with \eqref{eq:LambdaEquation} is now non-degenerate, hence we can express:
\begin{equation}
	\Lambda_{ij} = \sum_{ab} c_{ij}^{ab}( X^2 ) \mathcal{D}_{ab} + c_{ij}^{\Lambda}(X^2) \Lambda_{\text{total}}
\end{equation}

This means that we can now express the differential operators $\EE_{n_{ij}}$ in terms of $D_{ab}$ and $X^2_{ij}$, meaning that it can be Fourier transformed:
\begin{equation}
\EE_{n_{ij}}\left(X^2_{ij},|\Lambda_{ij} \right) \quad \longrightarrow \quad \hat{\mathbf{E}}_{ij}\left(s_{ab}, \dfrac{\p}{\p s_{ab}} \right)
\end{equation} 
where $s_{ab}$ is some choice of momentum space invariants. For example, by choosing the following parametrization:
\begin{equation}
	s_{ab}=2k^\mu_a k_{\mu,b}
\end{equation}
we get:
\begin{equation}
	\begin{split}
				\tilde{\mathcal{D}}_{ab} F\left( \{ s_{ab} \} \right)&= \left(\dfrac{\partial}{\partial k_b^\mu} - \dfrac{\partial}{\partial k_a^\mu}\right)\left(k_a^\mu+k_b^\mu\right)  F\left( \{ s_{ab} \} \right) = \\
				&=\sum_{c} (s_{ab}+s_{bc}) \left(\dfrac{\partial}{\partial s_{bc}}- \dfrac{\partial}{\partial s_{ac}} \right) F\left( \{ s_{ab} \} \right)
	\end{split}
\end{equation}

\subsection{Example of triangle}
The position space equation for a three point triangle function are written in terms of three operators $\Lambda_{12},\Lambda_{23},\Lambda_{12}$. According to our general prescription we can express them through operators, which are written in terms of the vertex positions $x_1^\mu,x_2^\mu,x_3^\mu$. To do this, we solve the following operator equations:
\begin{equation}
\begin{split}
        \Lambda = x_1^\mu \dfrac{\partial}{\partial x_1^\mu}+x_2^\mu \dfrac{\partial}{\partial x_2^\mu}+x_3^\mu \dfrac{\partial}{\partial x_3^\mu} &= \Lambda_{12}+\Lambda_{13}+\Lambda_{23}
        \\
        \mathcal{D}_{12}=(x_1^\mu-x_2^\mu)\left( \dfrac{\partial}{\partial x_1^\mu}+ \dfrac{\partial}{\partial x_2^\mu} \right)& = \dfrac{X_{12}^2+X_{13}^2-X_{23}^2}{2 X_{13}^2}\Lambda_{13} - \dfrac{X_{12}^2+X_{23}^2-X_{13}^2}{2 X_{23}^2}\Lambda_{23}
         \\
        \mathcal{D}_{13}=(x_1^\mu-x_3^\mu)\left( \dfrac{\partial}{\partial x_1^\mu}+ \dfrac{\partial}{\partial x_3^\mu} \right)& = \dfrac{X_{12}^2+X_{13}^2-X_{23}^2}{2 X_{12}^2}\Lambda_{12} - \dfrac{X_{13}^2+X_{23}^2-X_{12}^2}{2 X_{23}^2}\Lambda_{23}
\end{split}
\end{equation}
Solving this system we will get the matrix $M$:
\begin{equation}
\tiny{
\begin{pmatrix}
    X_{12}^2 (X_{12}^2 - X_{13}^2 - X_{23}^2) & -\dfrac{2 X_{12}^2 X_{13}^2 (X_{12}^2 - X_{13}^2 - X_{23}^2)}{X_{12}^2 + X_{13}^2 - X_{23}^2} & - \dfrac{2 X_{12}^2 (X_{12}^2 X_{13}^2 - X_{13}^4 + X_{12}^2 X_{23}^2 + 2 X_{13}^2 X_{23}^2 - X_{23}^4)}{X_{12}^2 + X_{13}^2 - X_{23}^2} \\
    & & \\
    -X_{13}^2 (X_{12}^2 - X_{13}^2 + X_{23}^2) & \dfrac{2 X_{13}^2 (X_{12}^4 - X_{12}^2 X_{13}^2 - 2 X_{12}^2 X_{23}^2 - X_{13}^2 X_{23}^2 + X_{23}^4)}{X_{12}^2 + X_{13}^2 - X_{23}^2} & \dfrac{2 X_{12}^2 X_{13}^2 (X_{12}^2 - X_{13}^2 + X_{23}^2)}{X_{12}^2 + X_{13}^2 - X_{23}^2} \\
    & & \\
    -X_{23}^2 (X_{12}^2 + X_{13}^2 - X_{23}^2) & 2 X_{13}^2 X_{23}^2 & 2 X_{12}^2 X_{23}^2 \\
\end{pmatrix}
}
\end{equation}
To decrease the size of matrix we omitted the common multiplier:
\begin{equation}
    \text{den}=\dfrac{1}{X_{12}^4 - 2 X_{12}^2 X_{13}^2 + X_{13}^4 - 2 X_{12}^2 X_{23}^2 - 2 X_{13}^2 X_{23}^2 + X_{23}^4}
\end{equation}
Thus the solution is:
\begin{equation}
    \begin{pmatrix}
        \Lambda_{12} \\
        \Lambda_{13} \\
        \Lambda_{23}
    \end{pmatrix}
    = \text{den}\cdot M \cdot 
    \begin{pmatrix}
        \Lambda \\
        \mathcal{D}_{12} \\
        \mathcal{D}_{13}
    \end{pmatrix}
\end{equation}
It can be seen from the size of matrix, that even in the simplest case the Fourier transformation becomes rather challenging, what makes its practical usage, i.e. connecting simple coordinate equations with difficult momentum ones, obstructed.

\section{Relation to PF}
Comparing with the case of banana integrals studied in \cite{Mishnyakov:2024rmb, Mishnyakov:2023sly}, connecting coordinate and momentum equations here has additional complications. There are two new arising problems, namely:
\begin{enumerate}
    \item Instead of single equation we should ``factorize'' the system of equations.
    \item Previously we had single variable non-fuchsian equation, what already produced a lot of complications. However, in this case, every equation has derivatives in multiple variables.
\end{enumerate}
Summing up, because of these reasons Fourier transformed equations cannot be considered independently but as the system. Thus there is no factorization for single equations and it is replaced by checking that the differential ideal generated by Fourier transformed system lies in the ideal generated by Picard-Fuchs system. In other words, the Fourier system should be the consequence of Picard-Fuchs one. Unfortunately, the complexity of these systems does not allow to present ``factorization'' rules, even in massless case, as we have done for equal mass case for banana diagrams in \cite{Mishnyakov:2024rmb}.

\section{Conclusion}

In this short letter, we have demonstrated that differential equations \emph{exist for any Feynman integral} and have a very transparent origin. The two main principles that we utilize is to work with \emph{unintegrated} functions and work in \emph{position space} initially. Let us summarize the main advantages of our approach.
\begin{itemize}
	\item  We have shown that existence and even shape of equations in position space can be determined from first principle. One can immediately trace its origin to just the equations of motion, i.e. the Gaussian part of the action.
	\item  The existence of equations in momentum space is an immediate corollary and  is in principle defined by explicit Fourier transform rules.
	\item In fact the discussion did not really depend that much on the shape of the e.o.m.. In principle, one could substitute the Klein-Gordon equation by any differential operator (perhaps of finite order) (see \cite{Bazarov:2023kgo} for a similar treatment in quite a different setting). Thus, the position space equations exist in any theory (defined perturbative through a quadratic action). 
	\item The position space equations can by potentially defined in cured space, to deal with loop correction in curved space QFT \cite{Cacciatori:2024zbe,Cacciatori:2024zrv,Chowdhury:2023arc}. 
\end{itemize}
The last point also prompts us to discuss another feature of the proposal. A particular part of the discussion of the Fourier transform is the possibility to integrate over certain vertices by setting some momenta in Fourier space to zero. However, the operation of an integration over a vertex can perhaps be archived directly in position space on the level of equations. The idea here is in the spirit of D-module theory. There one can rigorously define the so-called integration ideal in the D-module, and its dual restriction ideal in Fourier space. Effectively it gives a way to define differential equations that are satisfied by an integral of a function over some subset of the variables. In this context one requires certain holonomic properties from the functions and D-modules in question, which have to be investigated in our context. An additional complication that we deal with here is that the integration is over multi-dimensional vertex coordinates, while the differential equations are initially given in terms of invariants. Here again, the discussion in sec. \ref{sec:ChoiceOfVariables} is of relevance. Despite that we can still speculate that after appropriate modification techniques of D-modules and holonomic functions theory can be applied in our context exactly in the desired way: obtaining differential equation for integrated diagrams from the unintegrated ones. Concluding this discussion we mention that that would be the supposed approach to curved space integrals, where, in general, only position space methods are available.

\section*{Acknowledgements}
Nordita is supported in part by NordForsk. The work 
is supported by the Russian Science Foundation (Grant No.20-71-10073).

\bibliographystyle{utphys}
\bibliography{FeynmanRef}{}
\appendix 

\section{The banana case and review of the $\Lambda$-formalism}\label{sec:Appendix1}

In \cite{Mishnyakov:2023sly} a method of computing coordinate--space differential equations was proposed, which are essentially tensor powers of the equations of motion. Here we review some parts this method, that produces the differential operators $\EE_n(X)$.   All the operators are written using $\Lambda:=x^\mu  \p_\mu$.Then each propagator satisfies 
$$(\Lambda^2 + (D-2)\Lambda + X^2 m^2)G_m (X) = 0$$.
The banana function the product of propagators:
\begin{equation}
    B_n(\mathbf{m}|X)=\prod\limits_{i=1}^{n} G_{m_i}(X)
\end{equation}.
Firstly, we notice that one can reduce any power of the dilatation operator acting on the propagator to first derivatives and multiplication by $X^2$ using the equations of motion. Namely:
\begin{equation}\label{eq:LambdaOnG}
	\Lambda^k G_m =  \sum_{i=0}^{k} a_{k,i} X^{2i} \Lambda G_m + \sum_{i=1}^k b_{k,i} X^{2i}  G_m
\end{equation}
There does not seem to be any simple expression for these coefficients, however they can be defined recursively. Next, let us introduce a notation for a product of $\Lambda$-derivatives of the propagators:
\begin{equation}
	I_{\vec{k}}=I_{\left(k_1,\ldots k_n \right)}:= \prod_{i=1}^{n} \Lambda^{k_i} G_{m_i}
\end{equation}
where $k_i$ can be zero. In these notations the initial banana function is just:
\begin{equation}
	B_n(x)=I_{\vec{0}}:= I_{\left(0,\ldots \right)}
\end{equation}
Thus, applying the dilatation operator to the product of propagators we get:
\begin{equation}\label{eq:LambdaOnI1}
	\Lambda^n I_{\vec{0}} = \sum_{\vec k : \sum k_i=n} \binom{n}{k_1,\ldots k_n} I_{\vec{k}}
\end{equation}
Using the expansion \eqref{eq:LambdaOnG} for each of the $I_{\vec{k}}$ functions we obtain:
\begin{equation}\label{eq:LambdaOnI2}
	\Lambda^k I_{\vec{0}}=  \sum_{\vec{\epsilon}\,:\,\epsilon_i={0,1} } \!\! C_{k}^{\vec{\epsilon}} I_{\vec{\epsilon}}
\end{equation}
where the coefficients $C_{k}^{\vec{\epsilon}}$ are made up of the $a_{k,i},b_{k,i}$  of \eqref{eq:LambdaOnG} and the multinomial coefficients in \eqref{eq:LambdaOnI1}.
\\

The key idea is to notice that these relations for $k=1,\ldots 2^n$  imply that the homogeneous linear system with matrix elements $ \left(C_{k}^{\vec{\epsilon}} | \Lambda^{k} I_{\vec{0}}  \right)$ has a nontrivial solution given by the $2^{n}+1$--tuple $\left( \left\{I_{\vec{\epsilon}} \right\}_{\epsilon_i={0,1}} , -1 \right)$
This means that the determinant of this matrix should vanish which implies a differential equation on $I_{\vec{0}}=B_n$:
\begin{equation}\label{eq:DeterminantEq}
	\mathcal{E}^{(n)}_x I_{\vec{0}}:=\det_{\substack{k = 0,\ldots, 2^n \\ \vec{\epsilon} \in \mathbb{Z}_2^n }} \left( C_{k}^{\vec{\epsilon}} | \Lambda^{k} I_{\vec{0}}  \right)  = 0
\end{equation}

The coefficients $C_{k}^{\vec{\epsilon}}$ are in principle computed via the recursion for and $a_{k,i},b_{k,i}$, however, there is no explicit formula for them in terms of the masses and dimensions. Clearly, the final determinant is a more complicated object. However, formula \eqref{eq:DeterminantEq} establishes the existence and the form of the position space equations for generic $n$-banana functions with distinct masses. 
\\

To illustrate the construction, consider the example of $n=2$ generic mass banana diagram. In this case, the basic function is $I_{0,0}=B_2 = G_{m_1} G_{m_2} := G_1 G_2 $ and there are three ways to apply the dilatation operators:
\[ I_{1,0} = (\Lambda G_1) G_2;\ I_{0,1} = G_1 (\Lambda G_2);\ I_{1,1} = (\Lambda G_1) (\Lambda G_2). \] To get an equation we need to apply derivatives to the banana function four times. By iterated application of $\Lambda$ operator and using the relations \eqref{eq:LambdaOnG} we obtain:
\begin{equation}\label{eq:n2example}
	\begin{split}
		& \phantom{{}^0}\Lambda I_{0,0} = I_{1,0} + I_{0,1} \\
		& \Lambda^2 I_{0,0} = 2 I_{1,1} -(D-2)I_{1,0} - (D-2) I_{0,1} - (m_1^2 + m_2^2 )x^2 I_{0,0}  \\
		& \Lambda^3 I_{0,0} = -6 (D-2)I_{1,1} - (x^2 m_1^2  + 3 X^2 m_2^2 - (D-2)^2)I_{1,0} -
		\\
		& \phantom{\Lambda^4 I_{0,0}=} - (3 X^2 m_1^2 + m_2^2 X^2 -(D-2)^2)I_{0,1} + (D-4) (m_1^2 + m_2^2) X^2 I_{0,0} \\
		& \Lambda^4 I_{0,0} = (-4 (m_1^2+m_2^2) X^2+14 (D-2)^2) I_{1,1} +(2 X^2 ((D-4)m_1^2 + (5D-14)m_2^2) -(D-2)^3)I_{1,0} +\\
		& \phantom{\Lambda^4 I_{0,0}=}+(2 X^2 ((5D-14) m_1^2+(D-4) m_2^2)-(D-2)^3) I_{0,1} +\\
		& \phantom{\Lambda^4 I_{0,0}=} (-(m_1^2+m_2^2)(D^2 -6 D +12)+ 6
		m_2^2 m_1^2 X^2+m_1^4 X^2+m_2^4 X^2) I_{0,0} \\
	\end{split}
\end{equation}
The matrix $\left(C_{k}^{\vec{\epsilon}} | \Lambda^{k} I_{\vec{0}}  \right)$ is then explicitly given by:
{\small
\[ \begin{pmatrix}
	1 & 0 & 0 & 0 & I_{0,0} \\
	0 & 1 & 1 & 0 & \Lambda I_{0,0} \\
	- (m_1^2 + m_2^2 )x^2 & -(D-2) & -(D-2) & 2 & \Lambda^2 I_{0,0} \\
	(D-4)(m_1^2 + m_2^2)x^2 & - (m_1^2 X^2 + 3 m_2^2 X^2 -(D-2)^2) & -(3 m_1^2 X^2 + m_2^2 X^2 - (D-2)^2) & -6(D-2) & \Lambda^3 I_{0,0} \\
	C_4^{00} & C_4^{10} & C_4^{01} & C_4^{11} & \Lambda^4 I_{0,0} \\
\end{pmatrix}\]
}
where
\begin{equation}
	\begin{split}
		C_4^{00} &=  X^2 (-(m_1^2+m_2^2)(D^2 -6 D +12)+ 6
		m_2^2 m_1^2 X^2+m_1^4 X^2+m_2^4 X^2),	 \\
		C_4^{10} &= (2 X^2 ((D-4)m_1^2 + (5D-14)m_2^2) -(D-2)^3),  \\
		C_4^{01} &=(2 X^2 ((5D-14) m_1^2+(D-4) m_2^2)-(D-2)^3),  \\
		C_4^{11} &= -4 (m_1^2+m_2^2) X^2+14 (D-2)^2
	\end{split}
\end{equation}
The equations \eqref{eq:n2example} imply that the matrix has a kernel of the form $(I_{0,0}, I_{1,0}, I_{0,1}, I_{1,1},-1)$.
Therefore, its determinant vanishes, which produces an equation:
\begin{equation} \label{1loopcoord}
	\begin{aligned}
		\mathcal{E}^{(2)}_x I_{\vec{0}} =& 4 X^2 (m_1^2 - m_2^2) \Big\{ \Lambda^4 + 2 (2 D - 5) \Lambda^3  +
		\Big(2 X^2 (m_1^2 + m_2^2)+(D-2)(5D-16)\Big) \Lambda^2  +
		\\
		&+ 2 \Big((m_1^2+m_2^2) X^2 (2D-3)+(D-4) (D-2)^2 \Big) \Lambda  + \\
        &+ X^2 \Big((m_1^2 -m_2^2)^2 X^2 + 2 (D-1)(D-2) (m_1^2 +m_2^2)\Big) \Big\} I_{0,0} = 0
	\end{aligned}
\end{equation}

As we see, the equation is of order $4$ in $x$ derivatives just as expected. 
\subsection{Vanishing masses
		}
		Taking the limit of vanishing masses greatly simplifies the calculations. It is to be considered a ''solvable'' example: we can completely derive the form of the general position space operator starting from the equations of motion.  In this case the cut propagator satisfies:
		\be
		\Big(\Lambda^2 + (D-2)\Lambda\Big)G = 0
		\ee
		which makes applying powers of $\Lambda$ trivial:
		\begin{equation}
			\Lambda^{n}G = (2-D)^{n-1} \Lambda G \, , n>1
		\end{equation}
		Which makes the whole calculation much simpler. As a result we obtain the equation to be a determinant, analogous to the one in \eqref{eq:DeterminantEq}: 
		\begin{equation}
			\EE_n(X) = \left( \dfrac{n!}{(n-i)!} \right) \det C_n^{(\text{red})}
		\end{equation}
        where the reduced matrix is given by:
		\begin{equation}
			C_n^{(\text{red})}=\left\{\left( S_2(i,k) (2-D)^{i-k}| \Lambda^{i} \right) \right\}_{\substack{i=1,\ldots, n+1 \\
					k=1,\ldots ,  n
			} }
		\end{equation}
  where $S_2(i,k)$ are the Stirling numbers of the second kind. Computing the determinant gives:
  	\begin{equation}
			\EE_n(X)  =  \left(\prod_{i=0}^{n} \dfrac{n!}{(n-i)!} \right) \left(\prod_{i=0}^{n}(\Lambda+i(D-2))\right)
		\end{equation}
  As before vanishing of the determinant gives a differential equation:
		\be
		\EE_n(X)  B_n(X) =0
		\ee
For example:
At $n=2$ the derivatives are given by
			\begin{equation}
				\begin{split}
					\Lambda  G^2 &= 2G\Lambda G \nn \\
					\Lambda^2 G^2 &= 2G\Lambda^2 G + 2(\Lambda G)^2 = -2(D-2)G\Lambda G + 2(\Lambda G)^2 \nn \\
					\Lambda^3 G^2 &= -2(D-2)G\Lambda^2 G - 2(D-2)(\Lambda G)^2 +4\Lambda G \Lambda^2 G =
					2(D-2)^2 G \Lambda G - 6(D-2)(\Lambda G)^2
				\end{split}
			\end{equation}
			And the matrix:
			\begin{equation}
				M_2=\left(\begin{array}{ccc}
					2 & 0 & \Lambda Z_2 \\
					-2(D-2) & 2 & \Lambda^2 Z_2 \\
					2(D-2)^2 & -6(D-2) & \Lambda^3 Z_2 \\
				\end{array}\right)
			\end{equation}
			Hence the determinant gives
			\be
			\mathcal{E}_x^{(2)} = -4\Lambda\Big(\Lambda^2 + 3(D-2)\Lambda +2(D-2)^2\Big)
			= -4\Lambda \Big(\Lambda+(D-2)\Big) \Big(\Lambda+ 2(D-2)\Big)
			\label{vanishing_1_loop_coord}
			\ee

\end{document}